\documentclass[aps,preprint,a4paper]{revtex4}%
\usepackage{amsfonts}
\usepackage{amsmath}
\usepackage{amssymb}
\usepackage{graphicx}%
\providecommand{\U}[1]{\protect\rule{.1in}{.1in}}
\begin{document}
\title{Time and Quantum Clocks: a review of recent developments }
\author{M. Basil Altaie}
\affiliation{Department of Physics, Yarmouk University, 21163 Irbid, Jordan.}
\author{Daniel Hodgson and Almut Beige}
\affiliation{The School of Physics and Astronomy, University of Leeds, Leeds LS2 9JT,
United Kingdom.}
\pacs{PACS number}

\begin{abstract}
In this review we present the problem of time in quantum physics, including a
short history of the problem and the known objections about considering time
as a quantum observable. The need to deal with time as an observable is
elaborated through some unresolved problems. The lack of a 
consistent theory of time is currently hindering the formulation of a
full-fledged theory of quantum gravity. It is argued that the proposal set forth by 
several
authors of considering an intrinsic measurement of quantum time, besides
having the conventional external time, is compelling. Recently several
suggestions have been put forward to revive the proposal of Page and Wootters
(1983), elaborating and resolving some of the main ambiguities of the original
proposal and opening new scope for understanding its content. The approach
followed in these new contributions exposes the need to go beyond the
limitations enforced by the conventional approach of quantum physics. The
attitude of covariant loop quantum gravity, in which it is called to
completely ignore time, is also discussed. This review could be a step forward
in an endeavour to reform our outlook of the unification of the theory of
relativity and quantum physics by furnishing the conceptual ground needed for
this goal. Intentionally, some technical details are avoided since we aim
to present the  approaches to resolve the problem in a simple way with the clearest
possible outlook. These can be looked up in the original references 
provided.
\end{abstract}
\date[date: 15 March 2022]{}
\startpage{1}
\endpage{35}
\maketitle

\section{Introduction}

The concept of time has long been the subject of philosophical contemplation
as well as physical judgement for its role in defining the dynamics of
systems. For Aristotle, time was a measure of motion: it is a number of change
\cite{Arst}. If there is no change there is no time; that is to say time exist
whenever events exist. Even if there is no external change, the psychological
time for Aristotle is a form of change within ourselves. This makes space,
motion and matter constitute the arena in which time is realized. In this view
the universe was assumed to be eternal, with no beginning, and therefore time
was thought to always exist. The Algerian Christian theologian and philosopher
St.~Augustine of Hippo (356-430 C.E.) presented a different notion of time,
adopting the assumption that the universe originally had a beginning. Although
he considered time in relation with events as it was in the Aristotelean
conception, he confessed that time did not exist before the creation of the
universe. A beautiful and meaningful explanation for eternity is given in his
book \textquotedblleft confessions\textquotedblright\cite{Aug}. Abu Hamid
Al-Ghazali (1058-1111 C.E.), the Muslim theologian and philosopher, elaborated
further on the notion of time being related to the occurrence of events. Like
Augustine, he believed that the universe had a beginning with time.
Furthermore, Al-Ghazali called to deal with space and time on equal footing,
and recognized that the `time extension', as he called it, must be treated on
an equal footing as the space extension. In his own words: \textquotedblleft
Similarly, it will be said that just as spatial extension is a concomitant of
body, temporal extension is a concomitant of motion.\textquotedblright%
\cite{Gh}. Time according to Al-Ghazali can only be recognized in
relationships to other events; there is no \emph{before} or \emph{after}
except relative to another reference. He says: \textquotedblleft There is no
difference between temporal extension that in relation [to us] divides
verbally into \textit{before} and \textit{after} and spatial extension that in
relation [to us] divides into \textit{above} and \textit{below}. If, then, it
is legitimate to affirm an \textit{above} that has no above, it is legitimate
to affirm a \textit{before} that has no real before, except an estimative
imaginary [one] as with the \textit{above}. Clearly, no
absolute time existed according to these two philosophers.

Isaac Newton adopted a completely different concept of time in which time is
considered as an `absolute flow' that exists independent of the events and the
observers; time is always there, marking the history of events. Newton
recognized that a kind of `true time' exists that passes independently of
things and of their changes \cite{New}. Determining the dynamics of motion and
change, the time $t$ is always measured by an external clock which has its
reference in the eternal astronomical motion. As it is paraphrased by the
Stanford Encyclopedia of Philosophy, Newton stipulated that \textquotedblleft
Absolute, true, and mathematical time, and from its own nature, passes equably
without relation to anything external, and thus without reference to any
change or way of measuring time (e.g., the hour, day, month, or
year).\textquotedblright\ \cite{Stan}. Time is uniform and endless with no
start nor an end, referred to by the ephemeris universal time. The
transformation laws that maintain invariance of the laws of Newtonian
mechanics are the Galilean transformations, which express the relationships
between the positions of events, while time is left to be an unanimous
parameter. The relative velocity of the frame of reference is the only
coupling factor between the different frames of reference. In a sophisticated
formulation, the dynamics is usually described by the Hamiltonian theory in
which time plays a fundamental role in the evolution of the system. Changes
that take place in accordance with the effects of the involved forces, and
within which space and time are independent variables, are well-defined by
this dynamics.

In the theory of relativity, different observers measure time differently,
depending on their relative velocity or the strength of the gravity at their
position: time dilates due to the relative speed of the frame of reference and
due to gravity, manifested by the gravitational redshift effect \cite{Wein}.
In special relativity both space and time are considered as variables of the
transformation laws describing the relationship between different frames of
reference. These are the Lorentz transformations which maintain the invariance
of the laws of electrodynamics. Both space and time are treated on equal
footing, and the only universal invariant is the spacetime interval
$ds^{2}=g_{\mu\upsilon}dx^{\mu}dx^{\upsilon}$ (sum over repeated indices
apply). For this reason and to maintain the invariance of the laws of physics,
the notion of `proper time' was established. The proper time is common to all
observers, since it is a co-moving measurement, like the wrist-watch time
which is the same for all.

In the theory of general relativity, gravity plays a crucial role in defining
time. Gravity is described by curved spacetime manifolds, and this description
has imposed a fundamental change on our view of time. As in special
relativity, time is relative and depends not only on the state of relative
motion of the observer, but on their position in the gravitational field. Two
different notions of time within the formalism of general relativity can be
distinguished. Specifically, we have the coordinate time $t$ that appears as
the argument of the field variable, for instance in $g_{\mu\upsilon}(x,t)$,
and the proper time $s$ measured along a given world line $W=W^{\mu}(\tau)$
parameterized by the variable $\tau$ and defined by \cite{Wein}%
\begin{equation}
s=\int\limits_{W}d\tau\sqrt{g_{\mu\upsilon}\frac{dW^{\mu}}{d\tau}%
\frac{dW^{\upsilon}}{d\tau}}\,. \label{Eq1}%
\end{equation}
The Einstein field equations are the equations of motion in the theory of
general relativity, which can be seen as second order evolution equations in
$t$. While the proper time maintains the covariance of the dynamics, the
coordinate time $t$ in these equations plays the same role as an evolution
parameter of the equations of motion and is the same as the ordinary
non-relativistic time. However, the physical interpretation of $t$ is very
different from the interpretation of the variable in the non-relativistic
theory. While non-relativistic time is the observable quantity measured (or
approximated) by physical clocks, relativity clocks measure, in general, the
proper time $s$ along their worldline, not $t$. The relativistic coordinate
$t$ is a freely chosen label with no direct physical interpretation. This is a
well-known consequence of the covariance of the Einstein equations under
general coordinate transformations. It is important to remind the reader that
the theory of relativity suggests that the universe is closed, and accordingly
time wise it is blocked.

In the standard formulation of quantum physics time is considered a parameter
measured by an external clock and is independent of the observer and the
system \cite{Pe}. This is the same as the Newtonian time. The role of time in
the canonical formulation of the quantum dynamics is like its role in the
classical Hamilton-Jacobi formulation. In Schr\"{o}dinger's formulation of
quantum physics, the dynamics of a system is described in terms of the
evolution of the state in time and follows the equation $|\psi(t)\rangle
=e^{-iHt/\hbar}|\psi(0)\rangle$ in which the Hamiltonian of the system is
driving the temporal evolution unitarily. Alternatively, the Heisenberg
equation of motion describes the evolution of the measurement in time where
$i\hbar\partial A(t)/\partial t$ $=\left[  A(t),H\right] $, here it is the
observable, more accurately the observation, which is evolving in time. In the
particular case that the observable commutes with the Hamiltonian, the system
will be represented by stationary states. James Hartle \cite{Hartle} has shown
that in both formulations there is a need for a preferred Newtonian time. In
the Schr\"{o}dinger formulation a preferred time enters centrally into the
formulation of the notion of the state and in the evolutionary laws of that
state, whereas in the Heisenberg formulation the need for preferred time
appears in order to define the ordering of the projections upon calculating
the conditioned probabilities of an event in multi-time measurements. This
requirement of preferred time for ordering the projections reflects the fact
that the Heisenberg formulation is describing states that do not evolve in
time, rather it is the process of observation which is evolving in time. For
this reason the projections in the Heisenberg formulation are not trivially
ordered as it is the case of the Schr\"{o}dinger picture. In comparison with
the sum-over-history formulation, Hartle finds that time does not enter the
formalism for computing the probabilities in such a central way. This finding
of Hartle is important for the realization of the role of time in both
pictures. We will see later how the situation changes on considering another
basic formulation and on adopting a different equation of motion.

But is the notion of time presented in the standard formulations of quantum
physics an inevitable property, or can there be an alternative? According to
Hartle \cite{Hartle1} the answer is negative. After thorough analysis of the
role of time in non-relativistic quantum cosmologies, in which part of the
system functions approximately as an ideal quantum clock, Hartle suggests that
our familiar notion of time in quantum physics is not an inevitable property
of a general quantum framework but an approximate feature of specific initial
conditions. This allows for an alternative formulation of quantum physics
taken in a wider scope in which time may play a different role.

The Stone-von Neumann uniqueness theorem implies that any pair of canonically
conjugate observables is essentially the canonical Schr\"{o}dinger pair,
i.e.~the position operator and the momentum operator \cite{HO}. In particular
both operators are necessarily unbounded from above and below. Then, from the
non-relativistic form of the Hamiltonian, it follows that if the potential $V$
is bounded from below, then the energy observable is also bounded from below.
Therefore, the energy observable cannot have a canonical conjugate observable
represented by a self-adjoint operator. This implies the nonexistence of a
self-adjoint operator representing the time observable. Furthermore, Wolfgang
Pauli \cite{P} argued against considering time as a dynamical variable. His
argument says that if the Hamiltonian is the generator of time translations as
suggested by the Schr\"{o}dinger equation, then this implies that any
reasonable definition of the time operator must be conjugate to the
Hamiltonian. Consequently, both time and energy must have the same spectrum
since conjugate operators are unitarily equivalent. Clearly this is not always
true; normal Hamiltonians have a lower bounded spectrum and often only have
discrete eigenvalues whereas we typically require that time can take any real
value from $-\infty$ to $+\infty$. Accordingly, Pauli concluded that
constructing a general time operator is impossible.

However, time measurements are quite common experimentally, therefore a
theoretical representation for them in quantum physics should exist. For this
reason and for many other arguments, several approaches to define quantum time
and quantum clocks to measure it were suggested. In this article we will
present the most prominent of the recent suggestions for quantum time and for
quantum clocks that can be used to measure it. The article is arranged as
follows: in Sec.~\ref{sec2} we present several prominent cases where the
definition and measurement of time stands as a problem. In Sec.~\ref{sec3} we
present the early proposals to construct a quantum clock. In Sec.~\ref{sec4}
we discuss the motivations to have a formulation for quantum time measurement
and, in this context, we discuss the need for adopting the Wheeler-DeWitt
constraint on the Hamiltonian. In Sec.~\ref{sec5} we present the recent
formulations for dealing with quantum time, both for continuous and discrete
energy scales. Meanwhile the resolution of the criticisms raised against the
early proposals are presented and assessed. Finally in Sec.~\ref{sec6} we
present a discussion of some open questions.

\section{Problems Involving Quantum Time}

\label{sec2}

The problem of time in quantum physics is related to several unresolved
questions and paradoxes. These include the question of how much time is spent
in tunnelling through a potential barrier in what is known as the Hartman
effect, the electronic transition time, the quantum Zeno effect and the travel
or arrival time. The reason for presenting these problems here is to show that
the problem of time in quantum physics is not exclusively related to quantum
gravity but is more general than that. These widely discussed controversial
issues are intimately connected with time measurements.

\subsection{Tunnelling Time}

Tunnelling through potential barriers is one of the most stunning and
non-intuitive phenomenon predicted by quantum physics, and physicists have
been calculating and measuring tunnelling times for a long time now. In 1962,
Hartman \cite{H} studied the dynamics of a wave packet tunnelling through a
rectangular potential barrier. He derived an analytical expression for the
time spent by a non-relativistic particle inside the barrier and found that it
is independent of the thickness of the barrier and less than the time required
for the particle to travel the same distance in free space. This phenomenon is
called the `Hartman effect' and has been verified experimentally many times,
using both photons and massive particles, under various circumstances and
employing different techniques \cite{HN,Ec etal,BD,
MR,Yang,Rob,SN,EN1,Stein,Spiel}. Many controversial interpretations are given
to these experimental results. Some authors suggest that the superluminal
speed implies a violation of causality \cite{EN1,EN2,N,Hi,AN,NA}. Others
conclude that the superluminal effect does not violate causality. The problem
is complicated further as it involves the question of time measurement and a
sizable amount of work was done to identify the correct definition and
measurement of the `arrival time'. For example, Winful
\cite{Win,Win1,Win2,Win3,Win4,Win5} negates the possibility of superluminal
speed through tunneling, arguing that the `apparent' superluminal speed is
caused by using certain confusing definitions for the transit time, proposing
that the group delay in tunneling is not a transit time but a lifetime and
hence should not be used to assign a speed of barrier traversal. Other authors
\cite{Low} argue that the wave packet spread in momentum space will be so
great as to invalidate the conventional spacetime description of the event;
accordingly they conclude that an actual measurement of an anomalously short
traversal time cannot be made.

Going through the most recent contributions regarding the time of tunneling
through potential barriers mentioned above, one finds that the problem is far
from being solved and from being fully understood. It is not even clear
whether the the tunneling particles are actually going through the potential
barrier or not. This question might be answered by an experiment using a
single shot with a particle that leaves a verifiable non-interacting trace
within the barrier. Generally, under certain considerations, one would expect
the particle to be at any time anywhere in the region surrounding the barrier.
It is only the measurement of the flux of particles in different regions that
selects which path the particle is following, a target which cannot be
achieved under quantum indeterminism. Consequently, one would expect that the
resolution of this problem may be found by considering the time of arrival in
the context of the quantum measurement of time. Foden and Stevens \cite{Fod}
considered calculating the tunneling time using a primitive quantum clock
coupled to the quantum system, but their calculations ran into difficulties.
The reason, we think, is due to coupling the clock in their calculation to the
quantum system, a situation which ought to lead to the inseparability of the
Hamiltonian, and therefore the Hilbert space in consideration. Similar
difficulties faced the work of Peres \cite{Per}, as we will see below, who
devised a clock coupled to the quantum system causing large disturbances of
the results.

\subsection{The Quantum Zeno effect}

If the wave function collapse interpretation is to be taken as an empirical
fact and not just a pictorial presentation, then a continuous measurement of
an excited state will prevent it from making a transition. This is known as
the `quantum Zeno effect' \cite{Sud}. As it is said, \textquotedblleft a
watched kettle never boils." This is another effect which has been predicted
theoretically and is related to the measurement of time. The effect is
explained in the conventional context of the Copenhagen school by the collapse
postulate, where it is understood that the continuous measurement of the
energy will cause the wave function to be in a situation under continuous
collapse to the same state preventing any transition from taking place. For a
detailed account of the quantum Zeno effect see for example the extensive
review by Facci and Pascazio \cite{Pao}.

Peres \cite{Pe1} considered the quantum Zeno effect (QZE)\ in the context of
studying the effect of the measuring apparatus on the dynamical properties of
the quantum system, particularly its decay law. He considered analyzing this
problem away from the argument based on quantum measurement theory. Peres
finds that under a very tight monitoring the decay is usually slowed down and
can even be halted. However, he also found that the Zeno paradox arises
directly from the Schr\"{o}dinger equation, which means that it is not a
result of measurement. This undermines the assumption of the wave function
collapse. However, Peres considered again this question in a second paper
\cite{Per} trying a resolution through measuring time by a quantum clock which
he proposed, as we will see later in Sec.~\ref{sec5}, affirming again that the
QZE is not an effect of measurement. Nevertheless, analyzing the results of
Itano {\em et al.}'s experiment \cite{Ita}, Beige and Hegerfeldt \cite{Beige} have
shown analytically that for a wide range of parameters, the short laser pulse
acts as an effective level measurement to which the usual projection postulate
applies with high accuracy, thus concluding that the projection postulate is
an excellent pragmatic tool for a quick and simple understanding of the
slow-down of time evolution in experiments of this type. They also recognize
that the corrections to the ideal reductions and their accumulation over $n$
pulses must be included, and a complete freezing does not seem possible
because of the finite duration of measurements. Indeed this is correct once we
know that the Zeno effect is exposed exclusively in measurements on a time
scale comparable to the Zeno time, as shown below.

Another question which arises in this context is the question of the duration
of a \textit{quantum jump}, a problem which has been considered by Cook
\cite{C} and later by Schulman \cite{Sch}. Analyzing the duration of a quantum
jump, Schulman obtained an expression for what he calls the `Zeno time,' which
is a measure for the duration of the measurement slowing the transition. It is
given by
\begin{equation}
t_{z}=\frac{\hbar}{\sqrt{\langle\psi|\left(  H-E_{\psi}\right)  ^{2}%
|\psi\rangle}}\label{S1}%
\end{equation}
where $E_{\psi}=\langle\psi|H|\psi\rangle$. The transition probability
calculated by Schulman is%
\begin{equation}
P_{interrupted}\simeq\exp\left(  -t\Delta t/t_{z}^{2}\right)  \,,\label{S2}%
\end{equation}
where $\Delta t$ is the duration between two successive measurements.
Obviously, as $t_{z}\rightarrow0$, the transition gets halted. In this context
time is taken to be a continuous, external and absolute coordinate as it is in
standard quantum physics. Here this might be well justified for events taking
place on atomic and molecular time scales, but when the Zeno time is as in
Eq.~(\ref{S1}) above, a quantum clock is needed to measure time, otherwise the
exponential decay law cannot be verified \cite{Pe1}.

\subsection{Arrival Time in Quantum Mechanics}

The detection of a particle usually happens once the particle interacts with
the detector, which is normally located at a finite distance from its source.
The moment of detection is called \emph{arrival time}. Theoreticians find that
the designation of an arrival time is a problematic part of the measurement in
quantum mechanics since it is related to the indeterminacy of the position and
momentum of the particle \cite{Gro,Leo,Gal}.

Perhaps Allcock \cite{All} was the first to present a rigorous proof that the
ordinary Hilbert space of a freely moving particle does not contain a set of
measurement eigenstates of the arrival time. This proof depends on extending
the frequency spectrum of the wave function to cover both positive and
negative values, while at the same time the spatial domain of the wave
function is halved. Delgado and Muga \cite{DM} (also see \cite{Mug} for an
updated account) find that a self-adjoint operator with dimensions of time can
be explicitly constructed, and it is shown that its complete and orthonormal
set of eigenstates can be used to define consistently a probability
distribution of the time of arrival at a spatial point. However, we should
note that this does not contradict the argument of Allcock about extending the
spectrum to cover the positive and negative energy range. All in all, the
formalism used by Delgado and Muga using the energy eigenstates $|E,\pm
\rangle$ and the momentum projection operator $\Theta(\pm\overset{\wedge}{P})$
along with the self-adjoint operator sgn$(\overset{\wedge}{P})=\Theta
(\overset{\wedge}{P})-\Theta(-\overset{\wedge}{P})$ confirms again the
essential need to cover the positive and the negative frequency ranges for the
solutions of the eigenvalue equation in order to circumvent the Pauli
objection. Here we might need to be careful about accounting for the negative
energy in terms of the negative frequency. Both may have effectively the same
contribution in some cases, but to generalize this might need further
investigation. As we will see below, a more advanced and less involved account
for the arrival time has recently been considered by authors using
calculations depending on a quantum clock which measures the intrinsic time.
This affirms again the necessity for a quantum clock.

\subsection{Time in Quantum Gravity}

Unifying quantum physics and the theory of relativity was the dream of Albert
Einstein and many others. In fact, as early as 1916 Einstein expected that
quantum theory will modify his theory of gravitation and is quoted saying
\textquotedblleft\ldots it appears that quantum theory would have to modify
not only electrodynamics, but also the new theory of
gravitation\textquotedblright\cite{E}. A theory which unifies quantum physics
and gravity\ would provide a comprehensive framework to explain the physical
phenomena on all scales, from the microscopic to the macroscopic level.
However, this goal faces great conceptual and technical difficulties. The
problem of time is one of the main obstacles in devising such a theory. One
facet of the problem is the conflicting concepts of time in quantum physics
and relativity theory, the conflict between the absolute and the relative, the
external and the local. For these reasons there are several approaches to
quantum gravity (for a comprehensive presentation of these approaches see
\cite{ORI}). The other facet of the problem arises in maintaining the
unitarity required by the main equation of motion in quantum physics, the
Schr\"{o}dinger equation.

Over the decades several approaches to quantum gravity were suggested,
many of those dropped short of achieving the goal and others are still
surviving. In two seminal reviews published in the early 1990's Isham
\cite{Isham} and Kuchar \cite{Kuc} formalized a deep analysis of the
discrepancies between the different approaches that were available at the
time. Here we are not considering such an analysis of the present approaches
since our main focus is on quantum time.

The problem of time in quantum gravity originates from the fundamental
conflict between the way the concept of time is used in quantum theory, and
the role it plays in a diffeomorphism-invariant theory like general
relativity. Many authors recognized that observable time can only be defined
intrinsically \cite{Bai, Dew,Whe,Mis,Yor,Whe1,Kuc2}. The affirmation of these
authors and the examples given in their cited articles confirm that having the
time operator corresponding to a quantum observable would require that time be
treated intrinsically. Consequently this has a fundamental implication on the
philosophy of accounting for a global system (called a universe) which
includes the clock and the quantum system under consideration. This would be
in agreement with the notion of the block universe implied by the theory of relativity.

Rovelli \cite{Rov} argues that the theory of general relativity has
fundamentally changed the way we deal with the dynamics of physical systems.
Instead of having a time that passes with the ticking of an external cosmic
clock, general relativity has adopted the \textit{relative} evolution of
observable quantities, not the evolution of quantities, as functions of time.
According to him this summarizes the problem of time for a quantum gravity
approach. After suggesting some technical alternatives, he concludes that
\textquotedblleft in order to build a quantum theory of gravity the most
effective strategy is to forget the notion of time all together, and to define
a quantum theory capable of predicting the possible correlations between
partial observables.\textquotedblright\ This argument is found in many of his
articles \cite{Rov2}. Along the same line of thought we find Julian Barbour
who presents the \textquotedblleft end of time\textquotedblright%
\ \cite{Bar}.\ At first this might sound fine, but in fact we soon find that
forgetting the notion of time altogether would make it difficult for us to
regain time as we go on to consider the real world as we conceive it
empirically. The statistical notion of thermodynamic time requires an ensemble
of states and this might not always be available.

Dynamics is all about changes under the field of force, and a change can only
be recognized with reference to two states, the one before and the one after.
However, in order to generalize tracking the change in other states, to
determine the dynamics involved, knowing how fast changes are occurring
relatively, we need to have a measure that can be applied to all the changes
taking place; such a measure should be a reference for comparison. This
measure, which is independent of all the states, is what is called the
intrinsic or internal time. This means that we cannot completely forget about
time, but perhaps we should assign it to an internal clock which measures the
change with reference to any two states and consider that as a basic measure
for the change. In this way we can have a reliable reference by which we can
regain the external coordinate time.

The quest for quantum gravity certainly needs some sort of quantization of
time for a quantum picture of the spacetime to become possible. Applications
of the quantum picture of time are not limited to the above, since it is
expected that the new formalism of discrete time will have serious
implications in areas of quantum computing and related subjects \cite{Boe}.

\subsection{Pauli's Objection Revisited}

Recently, several suggestions were made to circumvent Pauli's argument
against the possible construction of a general time operator \cite{LM,
Agu}. In many of these the negative energy states, which are a result of the
relativistic treatment, play a vital role. The wave function describing the
state of the system is a complex function of space and time, both of which are
considered in the Newtonian sense of being independent and absolute. For
example Aguillon \emph{et al.}~\cite{Agu} showed that Lorentz invariance,
Born's reciprocity invariance and the canonical quantization of special
relativity can provide, among other things, the existence of a self-adjoint
time operator that circumvents Pauli's objection. Leon and Maccone \cite{LM}
showed that Pauli's argument fails when we use an internal clock for time
measurement allowing for time to be represented by an operator which is
conjugate to the clock Hamiltonian, not the system's Hamiltonian. The total
Hamiltonian includes the clock and the system. However, we find that this
argument of the internal measurement of time, though providing a way to
consider a time operator conjugate to the clock Hamiltonian, is insufficient
to revoke Pauli's objection since the system Hamiltonian may stay bound from
below. The proper approach to overcome Pauli's objection is to have a
dynamical Hamiltonian which is unbound from below. The approach of Hodgson and
collaborators \cite{Hodg,Hodg2} on quantization of the electromagnetic field
in position space provides such a dynamical Hamiltonian and, therefore, may
provide a better argument to circumvent Pauli's objection within the context
of a more general theory of field quantization.

Moreover, Khorasani \cite{Kho} showed that Dirac's equation could be modified
to allow discrete time, an exact self-adjoint $4\times4$ relativistic time
operator for spin $1/2$ particles is found and the time eigenstates for the
non-relativistic case are then obtained. The result confirms that particles
can indeed occupy negative energy levels with vanishingly small but non-zero
probability, contrary to the general expectation from classical physics. This
work is a better confirmation of the possibility to have the Hamiltonian
unbounded from below since, for massive spin $1/2$ particles, the results
confirm the quantum mechanical speculation that particles can indeed occupy
negative energy levels with vanishingly small but non-zero probability,
contrary to the general expectation from classical physics. Accordingly,
Pauli's objection regarding the existence of a self-adjoint time operator is resolved.

\section{Early Proposals for Quantum Time}

\label{sec3}

In this section we present some important historical approaches to formulate
the quantum measurement of time using a quantum clock. Despite the fact that
these early proposals are outdated now, to present them here is important for
justifying the later proposals which have been put forward during the last
five years or so. The reader will observe the necessity of the diversion from
the old approaches and for this reason the shortcomings of those early
approaches are exposed. Nevertheless, the story of quantum time may not be
complete until we see the empirical outcome of the recent approaches.

\subsection{Wigner-Salecker proposal}

Wigner demonstrated that there are limitations of spacetime measurements due to the quantum
nature of test particles \cite{Wig}, and it was in this context that Salecker
and Wigner \cite{SW} introduced the idea of a quantum clock. They proposed to
use clocks only for measuring spacetime distances and avoided using measuring
rods which are essentially macro-physical objects. Their approach might be of
value for some quantum applications, however the Salecker and Wigner program
of quantum measurement of time went into difficulties as soon as the target
tuned out to find the minimum mass of the clock and the minimum uncertainty
involved in the measurement of time. An alternative less muddy consideration
would be to consider the measurement of probabilities.

\subsection{The Peres Clock}

Peres \cite{Per} proposed another method of constructing a clock telling
quantum time. At first Peres digitizes the time measurement by assuming a
number of angular momentum states $N$ designated by $j$ such that $N=2j+1$.
The states are prescribed by the wavefunctions
\begin{equation}
\phi_{m}(\theta)=(2\pi)^{-1/2}e^{im\theta},\text{ \ \ }0\leq\theta
<2\pi,\label{Eq2}%
\end{equation}
where $m=-j,....+j$. Another orthogonal basis is given by
\begin{equation}
v_{k}(\theta)=\frac{1}{\sqrt{N}}%
%TCIMACRO{\dsum _{m}}%
%BeginExpansion
{\displaystyle\sum_{m}}
%EndExpansion
e^{2\pi ikm/N}\phi_{m}(\theta).\label{Eq3}%
\end{equation}
For large $N$ we have a sharp peak at $\theta=2\pi k/N$. This can be
visualized as pointing to the $k^{th}$ hour with an angle uncertainty of
$\pm\pi/N$.

Peres defines the projection as $P_{k}v_{m}=\delta_{km}v_{m}$ and defines the
clock-time operator as $T_{c}=\tau%
%TCIMACRO{\dsum }%
%BeginExpansion
{\displaystyle\sum}
%EndExpansion
kP_{k},$ where $\tau$ is the time resolution of the clock (re-creation time).
Accordingly the eigenvectors of $T_{c}$ are $v_{k}$ and the eigenvalues are
taken as $t_{k}=k\tau$ with $k=0,...,N-1$ obtained as follows:%
\begin{equation}
T_{c}v_{k}=\tau%
%TCIMACRO{\dsum _{n}}%
%BeginExpansion
{\displaystyle\sum_{n}}
%EndExpansion
nP_{n}v_{k}=%
%TCIMACRO{\dsum _{n}}%
%BeginExpansion
{\displaystyle\sum_{n}}
%EndExpansion
n\delta_{kn}v_{k}=\tau kv_{k}=t_{k}v_{k}, \label{Eq4}%
\end{equation}
where the initial state of the clock is always taken as $v_{0}$. Now,
measuring the time by this clock, $T_{c},$ will give an approximate discrete
approximation to the true time, just like reading a digital watch. This means
that the system is defining its own time. This time is so defined by the
Hamiltonian causing the unitary development of the system.

Peres then introduces the Hamiltonian of the clock as $H_{c}=\omega J$ with
$J=-i\hbar\partial/\partial\theta$ and with the eigenvalues of the clock given
by $H_{c}\phi_{m}=m\hbar\omega\phi_{m}.$ This is then incorporated with the
system whence the total Hamiltonian becomes $H=H_{c}+H_{s}$. In this
formulation the effect of the measuring device is incorporated into the
quantum system and $[T_{c},H_{c}]=i\hbar.$ However, this will lead to
difficulties as we shall see below.

Peres' quantum clock would interfere with the system and therefore a quantum
clock with higher resolution, despite being more classical, will cause a
greater disturbance in the system. For this reason the clock measurements run
into difficulties upon considering some applications, which include the time
of flight (arrival time) where he finds that his clock can only deal with the
case $E \gg\left\vert V\right\vert $ upon considering the problem of barrier
penetration. While in the case of Schr\"{o}dinger's cat, viewed as a decay
time problem, and upon considering using the clock for measuring the spin
precession under controlled switching times, Peres finds that his clock
readings will be overwhelmed by its coupling to the device which it is
supposed to control.

The work of Peres taught us that the clock must not interact with the system,
otherwise the precession of the clock will be limited and a higher choice of
the precession will lead to clock-system interference which would limit the
usage of such a clock. However, the sentence spoken by Peres on the invalidity
of differential calculus when dealing with quantum physics is very important
and is calling for the use of a different mathematical construction to deal
with physical systems. In fact, the adopted equation of motion has an
important role to play too, as will be shown below. In Peres's words
\textquotedblleft It thus seems that the Schr\"{o}dinger wave function
$\psi(t)$... is an idealization rooted in classical theory. It is
operationally ill-defined (except in the limiting case of stationary states)
and should give way to a more complicated dynamical formalism, perhaps one
non-local in time.\textquotedblright However, we are not sure whether a more
complicated or less complicated dynamical formalism is needed, but certainly
further developments in this field show that a quantum clock adopting a
different understanding of the dynamics and a different philosophy of time
could offer a better alternative as we will see below.

Perhaps a better choice for an alternative mathematical construct for
quantum gravity is to consider non-commutative spacetimes. Since it is known
that different non-commutative spacetimes arise very naturally in most current
approaches to quantum gravity, Addazi \emph{et al.}~\cite{Add} published a
condensed and elegant review of such spacetimes exposing their main features
and citing related experimental analyses in connection with the gravitational
waves (GW) signal GW 150914 (see \cite{Kob,Bah}). Incidentally, such an
analysis of GW signals looking for the traces of non-commutative spacetimes is
necessary since effects from non-commutative spacetimes are expected to be
found in the GW spectrum as it was the case when QED effects were discovered
in the spectra of the atoms.

\subsection{The Page and Wootters approach}

\begin{figure}[t]
\centering
\includegraphics[width=3.5062in]{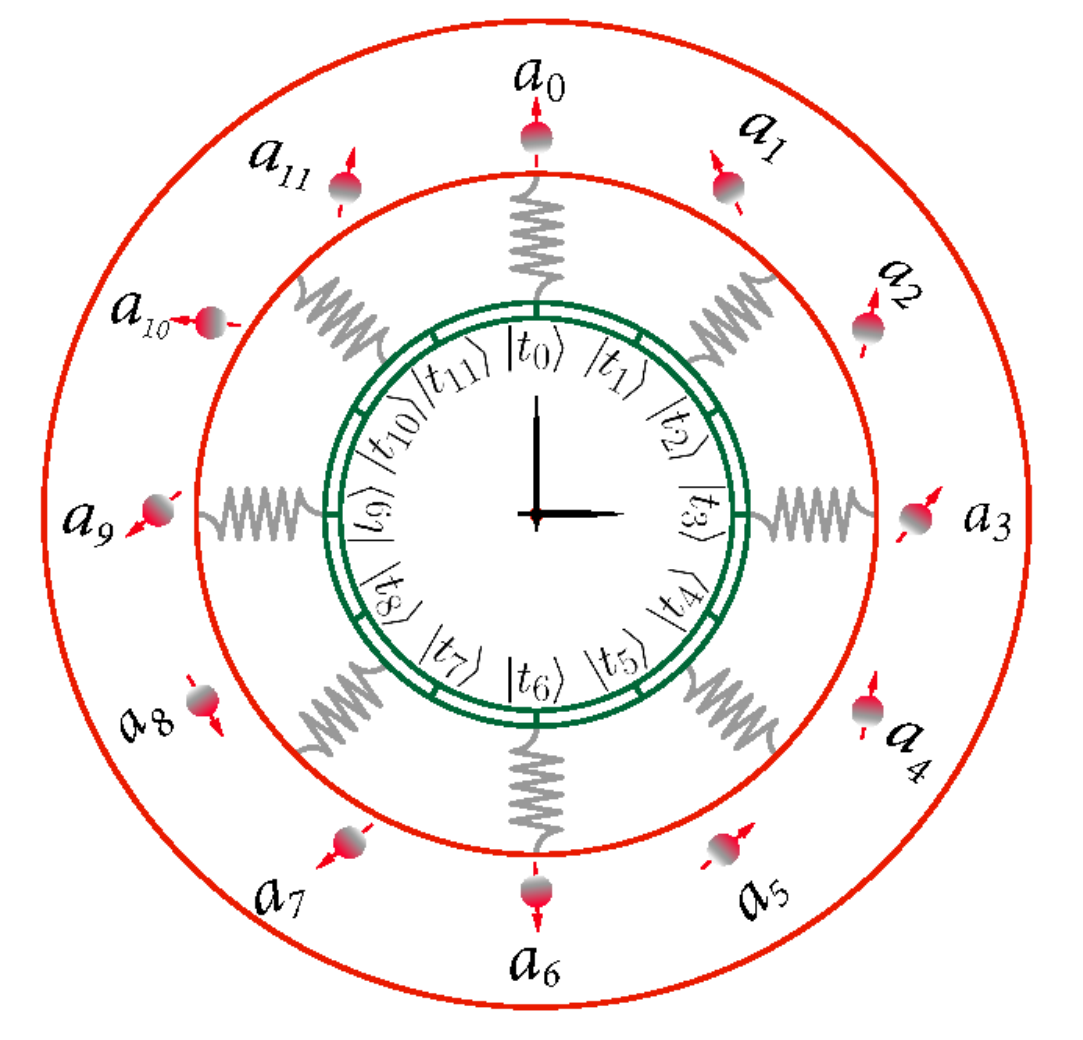}\caption{System observables
correlated with the clock readings.}%
\label{F0}%
\end{figure}

Page and Wootters (PW) \cite{PW} have placed several arguments and premises
for their proposal on constructing a quantum clock for measuring time. First,
influenced by the proof of Strocci and Wightman \cite{StW}, which showed that
the long-range Coulomb field causes the local charge operator to commute with
all quasi-local observables, they find that a superselection rule should exist
for gravitational energy. The observables in quantum systems should only be
those that commute with the Hamiltonian of the system. Accordingly, the
operator of time becomes stationary. This argument is placed to justify
dealing with the clock states as stationary ones and enables them to overcome,
at least partially, the problems faced by Peres when constructing a quantum
clock which led to his disappointment. They also argue that in the
Schr\"{o}dinger equation the time $t$ is an external parameter, and hence has
no place in the quantum theory if the system is truly closed. This argument is
placed to justify adopting the block universe proposal which the theory of
relativity is suggesting. Accordingly, they find that the temporal behaviour
of the observer in quantum systems depends on some internal clock time, not on
an external coordinate time. The clock system is decomposed into states of
definite clock times. The observed temporal behaviour of the closed quantum
system is found through the dependence of these component states on the time
labelling them. So we have the quantum clock measuring the internal time and
we have the values of the observables of the quantum system correlated with
those markings by the clock, see Fig.~\ref{F0}.

The proposal of PW has been criticized by several authors \cite{Kuc,Kuc1,Unr}.
Kuchar \cite{Kuc} argues that their quantum clock cannot be used to construct
a two-time propagator. In his view the PW quantum clock is unable to make two
moves, it will be frozen after the first move. To overcome this, Gambini
\textit{et al.}~\cite{Gam} used Rovelli's evolving constants of motion
\cite{Rov2}, parametrized by an arbitrary parameter which is then averaged
over to yield the correct propagators, thus saving the PW clock. Although the
end result matches the quantum predictions \cite{Mor}, Giovannetti \textit{et
al}.~\cite{G} find that the averaging which has been used by Gambini amounts
to a statistical averaging, which is usually reserved to unknown physical
degrees of freedom rather than to parameters with no physical significance.
So, the question of the two-time propagator in the PW approach remains open.
On the other hand, Unruh and Wald \cite{Unr} argue that the `conditional
probability interpretation' requires one to pick out a `preferred time
variable', or preferred class of such variables, from among the dynamical
variables. Furthermore, these authors seem to resort again to Pauli's
objection claiming that in Schr\"{o}dinger's quantum physics for a system with
a Hamiltonian bounded from below no dynamical variable can correlate
monotonically with the Schr\"{o}dinger time parameter $t$, and thus the role
of $t$ in the interpretation of Schr\"{o}dinger quantum physics cannot be
replaced by that of a dynamical variable. Obviously this is true for a
Hamiltonian which is bound from below, but the argument is ruled out in case
the Hamiltonian is unbound from both sides as will be shown below. Albrecht
and Iglesias \cite{And} raised some conceptual problems, since there are
several, inequivalent choices of the clock which then appear to produce an
ambiguity in the laws of physics for the rest of the universe: different
choices of the clock lead to different Hamiltonians, each corresponding to
radically different dynamics in the rest of the universe. So, it seems that
the logic of the timeless approach cannot be applied as directly as in
classical physics, because it does not lead to a unique Schr\"{o}dinger
equation for the rest of the universe. This clock ambiguity has also been
considered by Marletto and Vedral \cite{MV} and will be discussed below.

\section{Quantum Measurement of Time}

\label{sec4}

Recent presentations of the quantum time and constructions of quantum clocks
have adopted the view of a `block universe'. This construction is guaranteed
by considering the total Hamiltonian composed of two parts: the clock
Hamiltonian and the quantum system Hamiltonian. Accordingly, a joint Hilbert
space is formed out of the direct product of these two parts. Furthermore, a
Hamiltonian constraint is used to guarantee that the whole system forms a
block universe in which quantum states are stationary.

\subsection{Why do we need a quantum clock?}

There are two main reasons for that. First, we need to quantize time to have
it in harmony with the rest of the quantum structure. For this to be achieved
we need to deal with time as a quantum observable and not as a parameter.
Second, we need time to be an observer-dependent variable so that it would be
in harmony with the requirements of the theory of relativity, otherwise we
cannot obtain a construction for quantum gravity. For this goal to be achieved
we need to have time measured internally within the quantum system. The clock
must be separated from the quantum system, otherwise the quantum activity of
the clock may strongly interfere with the system's dynamics, causing high
disturbances as shown by Peres \cite{Per}. The quantum clock is a generator of
time eigenstates which are correlated with the quantum system's states.

\subsection{Why do we need a Hamiltonian constraint?}

Generally, Schr\"{o}dinger's equation does not satisfy the requirements
imposed by Einstein's relativity theory in respect to treating space and time
on an equal footing. Furthermore, relativity theory implies the notion of a
`block universe' where the past, present and the future all simultaneously
exist in the universe described by relativity theory. The \textit{before} and
the \textit{after} have no absolute status but are known relatively in a
relational description. This requires that the observables become stationary
in this universe. Such a requirement can be properly satisfied by applying a
constraint on quantum states in the form of an equation which is
sometimes called the Wheeler-DeWitt equation \cite{Page}%
\begin{equation}
H|\Psi\rangle=0. \label{q1a}%
\end{equation}
This equation is setting a constraint on the global state of the universe
(sometimes called the wavefunction of the universe). Its physical meaning may
be figured by saying that the global state $|\Psi\rangle$ (sometimes written
as double ket $|\Psi\rangle\rangle$) is a representation of a collection of
all stationary states contained in the universe (see Fig.~\ref{F1a}).

\section{Recent Developments}

\label{sec5}

The way in which PW presented their construction of quantum time was not as
clear as we have made it seem. Many questions and ambiguities were posed, and
for this reason perhaps the proposal did not attract enough attention for many
decades. In this section the PW approach is described for both continuous and
discrete spectra, as done in recent presentations. Presenting the PW formalism
in this way clarifies many apparent ambiguities and mysteries of the PW
approach. First, we briefly present the case for continuous spectra and then
offer more details for the case of discrete spectra as we find it more
suitable for the formulation of a discrete time measured by a quantum clock.

Recent studies of the quantum measurement of time and the assignment of a
quantum clock \cite{PW,G,MV,MS,FS} have adopted the constraint in
Eq.~(\ref{q1a}) as a basic prescription in order to assure the independence of
the clock which needs to behave as a subsystem belonging to a global system
represented by the state $|\Psi\rangle$. This means that the dynamics of the
conventional formulation of quantum physics is replaced by this constraint
equation where the total Hamiltonian $H$ is given by
\begin{equation}
H=H_{c}\otimes\mathbf{1+1\otimes}H_{s}. \label{q2a}%
\end{equation}
This Hamiltonian acts on the joint Hilbert space $\mathcal{H}_{c}%
\otimes\mathcal{H}_{s}$ which is a direct product of the two Hilbert spaces
$\mathcal{H}_{c}$ and $\mathcal{H}_{s}$, where $\mathcal{H}_{c}$ is spanned by
the basis states of the clock and where $\mathcal{H}_{s}$ is spanned by the
basis states of the quantum system. Taking $H_{c}=-i\hbar\frac{\partial
}{\partial t_{c}}$ as the clock Hamiltonian, one finds that
\begin{equation}
\left(  H_{s}+H_{c}\right)  |\Psi\rangle=\left(  H_{s}-i\hbar\frac{\partial
}{\partial t_{c}}\right)  |\Psi\rangle=0 . \label{q3a}%
\end{equation}
In this scheme the conventional formulation of quantum physics arises from
conditioning the reference time to $t$. This can be done by projecting the
global state $|\Psi\rangle$ onto a state $|t \rangle$ associated with the time $t$,
\begin{equation}
|\psi(t)\rangle=\langle t|\Psi\rangle , \label{q4a}%
\end{equation} 
which yields the state $|\psi(t)\rangle$ of the system at time $t$. 
The above is the basic structure of the formalism used in recent works for
devising a quantum clock which reads the time in correlation with the states
of the quantum system.

The solutions of Eq.~(\ref{q1a}) take different forms. Some authors describe
them in terms of a continuous formulation, whereas others adopt a discrete
formulation. The concern in these approaches is to get the probability of
happenstance of a state. This usually can be obtained using the Born rule with
the understanding that the calculated probability is conditional
\cite{PW,Page}. In other words, the happenstance of a given state of the
system is conditioned on the time $t$ in the Schr\"{o}dinger picture. (In the
Heisenberg picture the conditioning is on the observables.) Then a simple
Bayes conditioning of the Born rule probability of the joint state would allow
one to recover the full distribution of the time measurement \cite{MS}.
According to \cite{Page} the conditional probability $P(A|B)$ of a result $A$
given a testable condition $B$ is
\begin{equation}
P(A|B)=\frac{Tr(P_{A}P_{B}\rho P_{B})}{Tr(P_{B}\rho P_{B})}%
\end{equation}
where $\rho$ is the density matrix.

\subsection{Continuous spectra}

Giovannetti \textit{et al}. \cite{G}, and later Maccone and Sacha \cite{MS},
renovated the PW approach and elaborated the main features of this approach in
an elegant way, yet the basic arguments remain the same. Many problems faced
by previous proposals for measuring the time of arrival at a given position
are bypassed within their presentation. This enabled the authors to have a
generic calculation method that covers situations beyond the time of arrival
that other proposals could not treat \cite{Rus,Moga,Mug,Mie}. The particle
is described by a state $\ |\psi(t)\rangle$ with the time reference being a
continuous quantum degree of freedom in the Hilbert space $\mathcal{H}_{c}$ of
the clock. In this scheme we have a global state denoted by $|\Psi\rangle$
which represents all the states at once as explained above. This is now given
by
\begin{equation}
|\Psi\rangle=\frac{1}{\sqrt{T}}\int_{T}dt|t\rangle|\psi(t)\rangle. \label{G1}%
\end{equation}
Note here that the apparent entanglement in the above equation is not a result
of the clock-system dynamics, since both are isolated from each other as
required by the PW approach, but the result of having $|\Psi\rangle$
satisfying the constraint equation (\ref{q1a}). The states $|\psi(t)\rangle$
of the quantum system are obtained by projecting $|\Psi\rangle$ on $t;$ that
is $|\psi(t)\rangle=\langle t|\Psi\rangle$.

Extending the Born rule using the global state $|\Psi\rangle$ which sums up
the history of all the states of the system at all times, Maccone and Sacha
\cite{MS}\ choose to construct a positive operator-valued measurement (POVM)
of the time of arrival with the help of the projectors
\begin{equation}
\Pi_{t}=|t\rangle\langle t|\otimes P_{d} ~\ \text{and} ~ \Pi_{na}=1-\int
dt\Pi_{t}. \label{G2}%
\end{equation}
If the particle is at the detector $D$ then the projector is $P_{d}\equiv
\int_{D}dx|x\rangle\langle x|$. The projective POVM returns the value $t$ of
the clock if the particle is in $D$ or the value \textit{na} (not arrived) if
it is not, if the arrival observable $A$ is devised as
\begin{equation}
A=\int dt \, t \, |t\rangle\langle t|\otimes P_{d}+\mathbf{1\otimes\lambda
}\int_{x\notin D}dx|x\rangle\langle x| \label{G3}%
\end{equation}
where $\mathbf{\lambda}$ is an arbitrary eigenvalue of $A$ that signals that
the particle has not arrived. Using the Born rule, one obtains the joint
probability that the particle is at $x\in D$ and that the clock shows the time
of arrival is $t$ as%
\begin{equation}
p(t,x\in D)=Tr\left[  |\Psi\rangle\langle\Psi|\Pi_{t}\right]  =\frac{1}{T}%
\int_{x\in D}dx|\psi(x|t)|^{2}. \label{G4}%
\end{equation}
The time of arrival is recovered from the joint probability through the Bayes
rule as
\begin{equation}
p(t|x\in D)=\frac{\int_{x\in D}dx|\psi(x|t)|^{2}}{\int_{T}dt\int_{x\in
D}dx|\psi(x|t)|^{2}} \label{G5}%
\end{equation}
which is effectively averaging over $t$ through the whole period of the clock time.

\subsection{Discrete Spectra}

Discrete energy spectra need a slightly modified prescription. Pegg \cite{Peg}
recognized that the time representation of the clock's Hamiltonian with
$-i\hbar\frac{\partial}{\partial t_{c}}\rightarrow H_{c}$ would be correct
only if $H_{c}$ has a continuous unbound spectrum. In such a case the
Hermitian time operator in the energy representation can be written as
$\widehat{T}=-i\hbar\frac{\partial}{\partial E_{c}}.$ If one would consider an
isolated physical system of finite size then the introduction of an unbounded
Hamiltonian with continuous spectrum would not be possible. This implies that
a Hermitian time operator written in a differential form as above cannot be
introduced within the standard approaches according to Pauli's objection
\cite{P}. To overcome this dilemma, Pegg suggested constructing a Hermitian
operator named `Age' as a complement of a lower-bounded Hamiltonian with
equally spaced energy eigenvalues. His idea was to consider a Hamiltonian with
an energy cut-off, to calculate the quantities of interest and to eventually
remove the cut-off by allowing the upper bound to go to infinity. As such
`Age' becomes a generator of energy shifts while the Hamiltonian is the
generator of translation in time.

\begin{figure}[t]
\centering
\includegraphics[width=6.2124in]{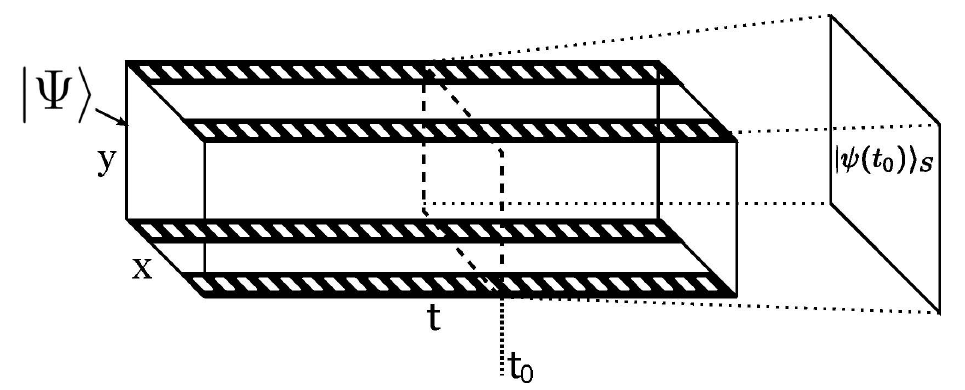}\caption{Pictorial representation of the
projection of the global state $|\Psi\rangle$ on a given moment of time
$t_{0}$ depicted here as a strip of a movie reel.}%
\label{F1a}%
\end{figure}

Favalli and Smerzi \cite{FS} introduced the prescription of Pegg into a PW
formalism and found that the `Age' can be interpreted as a proper Hermitian
time operator conjugate to a good clock Hamiltonian. With this amended
approach they could deal with equally spaced energy spectra as well as with
unequally-spaced spectra. Here we present this formalism since we feel that
its details present the case of quantum time measurement in a more suitable
way as both the energy representation and the clock readings are expressed in
discrete variables. Again, the Hamiltonian is described to contain two parts,
the system Hamiltonian $H_{s}$ and the clock Hamiltonian $H_{c}$ and is
structured as in Eq.~(\ref{q2a}) above. The system is constrained by the
Wheeler-DeWitt equation as in (\ref{q3a}) and the global wave function
$|\Psi\rangle$ for a discrete system is of the form
\begin{equation}
|\Psi\rangle=\frac{1}{\sqrt{T}}%
%TCIMACRO{\tsum _{n}}%
%BeginExpansion
{\textstyle\sum_{n}}
%EndExpansion
a_{n}|t_{n}\rangle|\psi(t_{n})\rangle.
\end{equation}
Here the quantum states of the system at time $t_{n}$ are represented by
$|\psi(t_{n})\rangle$. Furthermore, the wave
function $\psi(x|t_{n})$ can be obtained by projecting $|\psi(t_{n})\rangle$
on the position eigenbasis $\langle x|\psi(t_{n})\rangle$. A pictorial
representation of the projection of the states is shown in Fig.~\ref{F1a}.

\subsubsection{Eigenstates of the Clock}

Following Pegg's prescription \cite{Peg}, as elaborated by Favalli and Smerzi
\cite{FS}, a good clock is defined by a physical system governed by a
lower-bounded Hamiltonian with discrete equally-spaced energy levels (for
example, a simple quantum harmonic oscillator)
\begin{equation}
H_{c} = {\textstyle\sum_{n=0}^{s}} E_{n} |E_{n}\rangle_{cc} \langle
E_{n}|\label{Eq5}%
\end{equation}
where $d_{c} = s+1$ is the dimension of the Hilbert space $\mathcal{H}_{c}$ of
the clock and the $E_{n}$ and $|E_{n} \rangle_{c}$ are its energy eigenvalues
and eigenstates. The Hermitian time-operator of the system can then be defined
as
\begin{equation}
\widehat{\tau}=%
%TCIMACRO{\tsum _{m=0}^{s}}%
%BeginExpansion
{\textstyle\sum_{m=0}^{s}}
%EndExpansion
\tau_{m}|\tau_{m}\rangle_{cc} \langle\tau_{m}| \label{Eq6}%
\end{equation}
with the time eigenstates $|\tau_{m}\rangle$ given by
\begin{equation}
|\tau_{m}\rangle_{c} =\frac{1}{\sqrt{s+1}}%
%TCIMACRO{\tsum _{n=0}^{p}}%
%BeginExpansion
{\textstyle\sum_{n=0}^{p}}
%EndExpansion
e^{-iE_{n}\tau_{m}/\hbar}|E_{n}\rangle_{c} \label{Eq7}%
\end{equation}
with the orthogonality condition and completeness relationship
\begin{equation}
_{c} \langle\tau_{m}|\tau_{m^{\prime}}\rangle_{c} =\delta_{mm^{\prime}} ~~
\mathrm{and} ~~ {\textstyle\sum_{m=0}^{s}} |\tau_{m}\rangle_{cc} \langle
\tau_{m}|=\boldsymbol{1}_{c}. \label{qC}%
\end{equation}
It is easy to show that the above clock Hamiltonian $H_{c}$ is a generator of
transitions in time (see \cite{Per,Peg,FS}) and hence conjugate to $H_{c}$.
The eigenvalues of $\widehat{\tau}$ of a clock with equally-spaced levels can
be written as
\begin{equation}
\tau_{m}=\tau_{0}+m\left(  \frac{T}{s+1}\right)  \label{Eq8}%
\end{equation}
where $T$ is the time it takes the clock to return to its initial state (full
cycle). For example, for $E_{n} = E_{0} + \hbar n \omega$ and
\begin{equation}
T= \frac{2\pi}{\omega} , \label{Eq9}%
\end{equation}
we automatically have $|\tau_{m=s+1}\rangle_{c}=|\tau_{m=0}\rangle_{c}$ which
is a property of the clock. Conversely, the smallest time interval measured by
the clock is
\begin{equation}
\Delta\tau=\frac{2\pi}{(s+1)\omega}. \label{Eq10}%
\end{equation}

\subsubsection{Dynamics}

The global Hilbert space of the universe (i.e.~of~the system and the clock) is
$\mathcal{H}=\mathcal{H}_{c}\otimes\mathcal{H}_{s}$ with $\mathcal{H}_{c}$
having the dimension $d_{c}=s+1$ and $\mathcal{H}_{s}$ having the dimension
$d_{s}=p+1.$ In order to cover the whole range of states of the system with
clock readings we should always have $d_{c}\gg d_{s}.$ A general state of the
whole system (the universe) can then be written as
\begin{equation}
|\Psi\rangle=%
%TCIMACRO{\tsum _{n=0}^{s}}%
%BeginExpansion
{\textstyle\sum_{n=0}^{s}}
%EndExpansion%
%TCIMACRO{\tsum _{k=0}^{p}}%
%BeginExpansion
{\textstyle\sum_{k=0}^{p}}
%EndExpansion
c_{n,k}|E_{n}\rangle_{c}\otimes|E_{k}\rangle_{s}\label{qD}%
\end{equation}
where the $|E_{k}\rangle_{s}$ denote the energy eigenstates of the system. On
imposing the condition $H|\Psi\rangle=0$ and under the assumption of having a
sufficiently dense clock spectrum one finds that the possible states
$|\Psi\rangle$ in the above equation reduce to states of the form
\begin{equation}
|\Psi\rangle=%
%TCIMACRO{\tsum _{k=0}^{p}}%
%BeginExpansion
{\textstyle\sum_{k=0}^{p}}
%EndExpansion
c_{k}|E=-E_{k}\rangle_{c}\otimes|E_{k}\rangle_{s},\label{qE}%
\end{equation}
where $%
%TCIMACRO{\tsum _{k}}%
%BeginExpansion
{\textstyle\sum_{k}}
%EndExpansion
|c_{k}|^{2}=1$. Using Eq.~(\ref{qC}), the global state of the universe can be
written as a superposition of direct products of clock and the system energy
eigenstates \cite{FS},%
\begin{equation}
|\Psi\rangle=\frac{1}{\sqrt{s+1}}%
%TCIMACRO{\tsum _{m=0}^{s}}%
%BeginExpansion
{\textstyle\sum_{m=0}^{s}}
%EndExpansion
|\tau_{m}\rangle_{c}\otimes|\psi_{m}\rangle_{s},\label{qF}%
\end{equation}
where the $|\psi_{m}\rangle_{s}$ are the states of the system related to the
global state $|\Psi\rangle$ by
\begin{equation}
|\psi_{m}\rangle_{s}=(s+1)\langle\tau_{m}|\Psi\rangle\label{qG}%
\end{equation}
which are the states of subsystem $S$ conditioned on the clock $C$ being in
$|\tau_{m}\rangle$. As such, the time development of $|\psi_{m}\rangle_{s}$
follows
\begin{equation}
|\psi_{m}\rangle_{s}=e^{-iH_{s}\left(  \tau_{m}-\tau_{0}\right)  /\hbar}%
|\psi_{0}\rangle_{s}.\label{qH}%
\end{equation}
Expressed in terms of the unitary operator $U_{s}(\tau_{m}-\tau_{0}%
)=e^{-iH_{s}\left(  \tau_{m}-\tau_{0}\right)  /\hbar}$, the state of the
universe can now be written as
\begin{equation}
|\Psi\rangle=\frac{1}{\sqrt{s+1}}%
%TCIMACRO{\tsum _{m=0}^{s}}%
%BeginExpansion
{\textstyle\sum_{m=0}^{s}}
%EndExpansion
|\tau_{m}\rangle_{c}\otimes U_{s}(\tau_{m}-\tau_{0})|\psi_{0}\rangle
_{s}.\label{qI}%
\end{equation}
Hence the conditional probability of obtaining outcome $\alpha$ for system $S$
when measuring the observable $A$ at a certain clock time $\tau_{m}$ is
\begin{equation}
P\left(  \alpha\text{ on }S|\tau_{m}\text{ on }C\right)  =|\langle\alpha
|U_{s}(\tau_{m}-\tau_{0})|\psi_{0}\rangle|^{2}\label{qJ}%
\end{equation}
according to the Born rule.

\subsubsection{Unequally Spaced Energy Levels}

\begin{figure}[t]
\centering
\includegraphics[width=4.9245in]{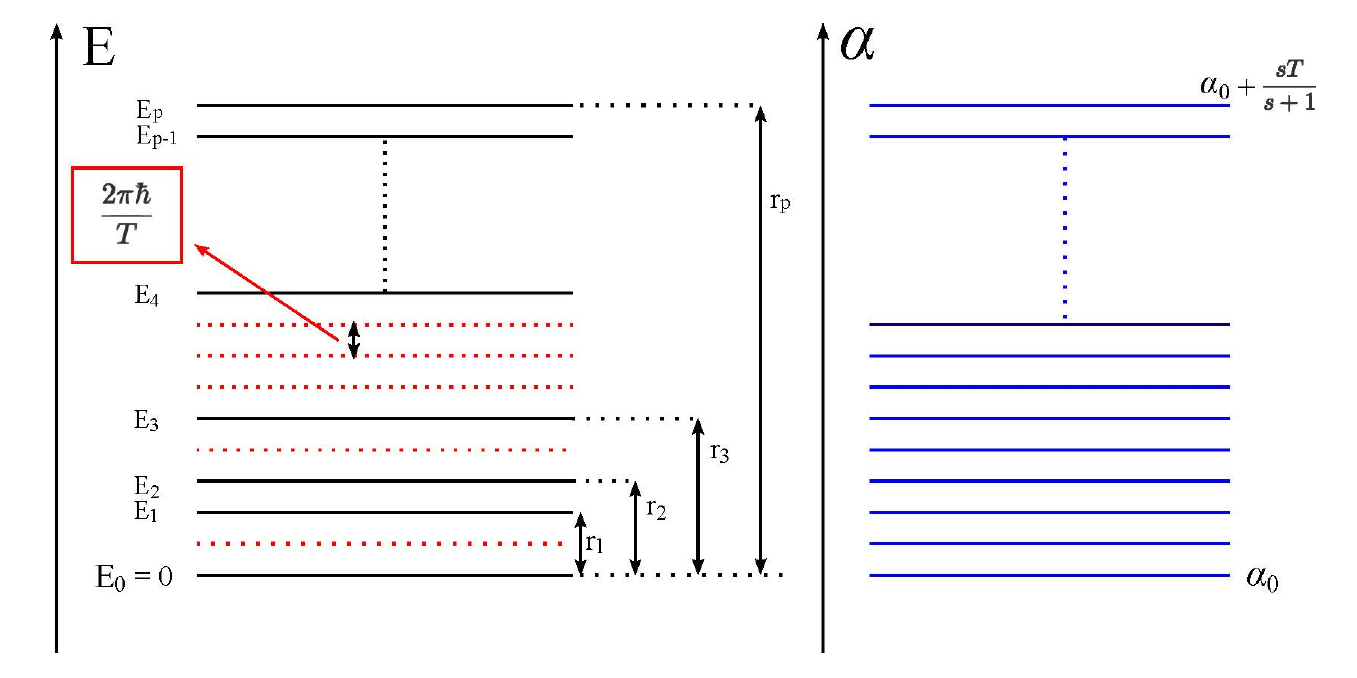}\caption{Shows the distribution of the
energy levels for unequal energy levels (left) and for equal ones (right).}%
\label{F2}%
\end{figure}

Suppose a quantum system has $p+1$ non-degenerate energy eigenstates
$|E_{i}\rangle_{s}$ with eigenenergies $E_{i}$ which are unequally-spaced but
have rational energy differences. These energies can be written as
\begin{align}
E_{i} =E_{0}+r_{i}\frac{C_{i}}{B_{i}} =E_{0}+r_{i}\frac{2\pi\hbar}{T}
\label{qK}%
\end{align}
where $C_i$ and $B_i$ are natural numbers,
as is the case of a free particle in a rigid box. Here $T = 2\pi r_{1}%
/(E_{1}-E_{0})$ and $r_{i}=r_{1} C_{i} / B_{i}$ with $r_{0}=0$. In such a case
it is not possible to define a Hermitian operator for time, but Pegg  
\cite{Peg} managed to identify a probability measure 
for such quantum systems
with quantum clock states that are defined in a similar fashion to the case of
equally-spaced energy levels. However, the rational ratio of the energy
differences now describes a ladder of time measurements with a distribution of
clock states along the unequally-spaced energy eigenstates of the system as
shown in Fig.~\ref{F2}. Using the clock states $|\alpha_{m}\rangle_{c}$ of an
associated quantum system with equal level spacing in a higher-dimensional
Hilbert space, the probability of obtaining an outcome $a$ for the system $S$
with unequal level spacing when measuring the observable $A$ at a certain time
$\alpha_{m}$ is again given by the Born rule as
\begin{equation}
P\left(  a\text{ on }S|\tau_{m}\text{ on }C\right)  =|\langle a|U_{s}%
(\alpha_{m}-\alpha_{0})|\psi_{0}\rangle|^{2} \label{qM}%
\end{equation}
for certain times $\tau_{m} = \alpha_{m}$ and for a sufficiently large
dimension $s+1$ of the clock and a sufficiently large $T$ \cite{Peg}.

\subsection{Clock Ambiguity}

Since quantum theory provides infinitely many non-equivalent ways of
partitioning the total Hilbert space of the universe into a tensor-product
structure, there are several choices of the clock by which unitary evolution
can arise in the rest of the universe. In other words, given the same global
state $|\Psi\rangle$ describing the universe, the PW approach leads to
completely different dynamics on the rest of the universe. This is the
\textit{clock ambiguity} which has been nicely presented and discussed
in\ \cite{Agu}. Clock ambiguity has also been thoroughly discussed by Marletto
and Vedral \cite{MV} who claim that there is no ambiguity in the PW formalism,
but there is something else worth elaborating here. They provide some
convincing arguments suggesting that the unitarity of the quantum system
safeguards the equivalence of the dynamics generated by different clocks
watching the same system. Indeed, their observation is analogous to the
equivalence of the inertial frames of reference in the theory of special
relativity. However, there is yet a need for a unifying reference for the
choices of different times generated through different partitionings of the
Hilbert space. Some attempts to resolve this question under linearized gravity
have been carried out by Giacomini \cite{Gia}, but their consideration of the
problem in a perturbative regime may undermine this attempt and turn it into
an exercise rather than a viable solution. Until a metric defining a
\textit{proper time} is formulated we cannot talk about the equivalence of
dynamics. This question remains an issue for further development of the
program of quantum time formulation.

\subsection{The Flow of Time}

In the PW approach the flow of time arises in the correlation (entanglement)
between the quantum degrees of freedom and the rest of the system, a
correlation that is present in the global, time-independent state
$|\Psi\rangle$. An internal observer will see such a state describing normal
time evolution; as we said earlier, the familiar system state $|\psi
(t)\rangle$ at time $t$ arises by conditioning $|\Psi\rangle$ via the
projection $\langle t|\Psi\rangle$. Effectively there is no flow of time in
the PW scheme. The clock measures the time correlated with the occurrence of
the states according to the given probability \cite{G}. This constructs a
history as illustrated in Fig.~\ref{F1a}, where the global state $|\Psi
\rangle$ is projected onto space while the dotted frame translates along the
horizontal axis representing $t$, just like the the unfolding of a movie reel.
The recorded movie frames are stationary states within a volume represented by
$|\Psi\rangle$ which constitutes a memory. Each frame stands for a given state
of the system and as the time selects a frame by a measurement of the clock it
gets projected onto a Schr\"{o}dinger state $|\psi(t)\rangle$.

Marletto and Vedral \cite{MV} describe this by saying that we have the
observer, the observed and a sequence of ancillas. The elaboration given by
Marletto and Vedral for the flow of time in the PW approach through counting
the sequence of ancillas is analogous to the translation in a space-like
universe. In such a universe, time is not time flowing independent of events
but closely linked to spatial landmarks through which we can construct a
sequence of correlations. From their own perspective, an observer does not
notice anything different in the direction of the sequence of the occurrence
of the events. Nevertheless, there is a meaning for the \textit{before} and
the \textit{after}, marked by the order of the state `before' and the state
`after.' So, there is a meaning for a change in a local sense. As correctly
remarked by \cite{MV}, this is because an observer does not notice any sense
of directivity, their state only contains information about the previous
times. Our remark here is on constructing a memory; it is hard to understand
how the observer can construct a memory in a universe of stationary states
except through different spatial locations. But if the spatial locations are
not known then there will be no memory except for the last location, the one
before the transition is made.

Hence, we conclude that the treatment of the transition from one state to the
next could perhaps be presented using Hidden Quantum Markov Models
\cite{Moras}. In such a treatment the full scope of transitions would be
presented as a correlation out of the set of possible states of the system and
the corresponding states of the clock. This again is an issue that is worth
further research. In a description based on Hidden Quantum Markov Models, the
order of time and the concept of causality would be altered to become a
correlation between cause and effect, thereby preserving causal order and
causal relationships; however, no causal determinism can be strictly
identified. This brings in the role of causal sets into the play of the
exposed dynamics. For example, Zych \emph{et al.}~\cite{Zych 1} considered
such a question in a time-like context.

\section{Discussion and Conclusion} \label{sec6}

The above approaches to the introduction of quantum time into the structure of
quantum physics are certainly important steps towards dealing with time as an
operator corresponding to a quantum observable, thereby targeting the role of
time in quantum gravity. The quantization of spacetime, once having considered
the bearing of the dynamics of events, is necessary for accurately describing
natural phenomena. Relativistic effects are observed only in some extreme
cases of high speed or strong gravity. While the dominating approach to
quantum gravity, namely the covariant loop quantum gravity, tends to ignore
the conventional spacetime background, the quantum time approach aims at
quantizing time by considering it an internal degree of freedom which is
observer dependent. In this respect the quantum time approach may face less
difficulties in unifying relativity theory with quantum physics since it would
be an difficult task to compile physical phenomena on the macroscopic level
without regard to space and time \cite{Ellis}. In another article
\cite{Ellis1}, which gives a beautiful account for when quantum effects are
significant, the picture of a classically Evolving Block Universe (EBU), in
his words, cedes place to one of a Crystallizing Block Universe (CBU) which
reflects this quantum transition from indeterminacy to certainty, while
nevertheless resembling the EBU on large enough scales. This issue is quite
worth consideration.

Starting from an overall quantum description of two entangled but
non-interacting systems, one of which is counted as a clock, Foti \emph{et
al.} \cite{Foti} take the classical limit of the clock and only obtain the
Schr{\"o}dinger equation in this limit. Upon taking the classical limit for
both the clock and the evolving system, they obtain Hamilton's equation of
motion. In their opinion, this shows that there is not a \textquotedblleft
quantum time\textquotedblright which is possibly opposed to a
\textquotedblleft classical\textquotedblright\ one; there is only one time and
it is a manifestation of entanglement. These results can be easily explained
by knowing that the time in Schr{\"o}dinger's equation is continuous and
consequently recovering the equation is expected. The Hamiltonian of the
global system is formed of the clock Hamiltonian and the system Hamiltonian,
and upon taking the classical limit of both systems, the clock Hamiltonian
gets dissolved and we are left with Hamilton's equation of motion. Indeed,
there are now two types of time, a discrete quantized time and a classical
continuous time. 

Considering a closed quantum system with a state that is perfectly
distinguishable from all past or future states, Stoica \cite{Sto,Sto2} shows
that for any change that happens the Hamiltonian must be $H=-i\hbar
\frac{\partial}{\partial\tau}$. Indeed, this is no surprise as this
Hamiltonian is expressing the presence of a quantum clock as in the
construction presented in this article \cite{Peg}, where $\tau$ is the time
identified with the clock time correlated with change in the limit of
vanishing energy level spacing. However, the value of this concise result
comes from its explanation of the the essence of the WD equation expressed as
a Hamiltonian constraint as in Eq.~(\ref{q3a}) and its connection with the
irreversibility of events. This is a result that might be taken further
within the context of describing the causal order in a block universe.

Notice that the constructed quantum clocks do not interact with the quantum
system for which the time is being measured. This is an important condition to
avoid the complications of having the clock disturbing the performance of the
system as shown by Peres \cite{Per}. This is true as long as the clock is not
physically related to the quantum system itself; however, the constraint of
the WD equation in (\ref{q3a})\ establishes the conjugation between the clock
and the system. The conditional probability is an important correlation link
between the system and the clock. Nevertheless, there still seems to be
several outstanding problems with the construction proposed by PW and other
variants that followed. To resolve such problems we may need to add a
fundamentally new postulate to the standard formulation of quantum physics
which should provide us with a comprehensive resolution of other basic
questions regarding the interpretation of quantum physics, like its range of
validity and the problem of quantum measurement in general. We need to
clarify, once and for all, the concepts we are dealing with. We need to
clarify the physical meaning of the WD equation and its important role in
fulfilling the requirement of the theory of relativity. We need to produce
some quantum effects in the relation between the clock and the quantum system
that is monitored. In this respect the question of entanglement needs to be
further studied since the consideration of Marletto and Vedral \cite{MV} is by
no means exhaustive. Moreover, we may need to add another postulate relating
the changes in the value of observables to the probability of occurrence of
the states, otherwise the clock may freeze after the first reading. These are
some of the open questions that remain to be answered by a full theory of
quantum time.

The identification of the WD constraint as a necessary prescription for the
dynamics of isolated quantum systems is certainly an important step towards
facilitating the adoption of the new philosophy of the relational role of time
in connecting stationary states. However, again this approach is by no means
complete. There seems to be more fundamental implications that need to be
recognized for the full picture to become plausible. Such implications get
uncovered, for example, in attempts of quantizing fields in position space. An
example is offered in \cite{Hodg,Hodg2}, where it is shown that the full range
of positive and negative frequencies is needed to represent electrodynamics.
The role of the negative frequency solutions is very important and, as the
solutions of the Dirac equation for the electron has shown, the treatment of
space and time on equal footing as required by the theory of relativity has
various implications that cannot be ignored. The existence of anti-particles
and the negative energy states in vacuum is certainly strong evidence of such
contributions brought out by the Dirac equation.

If the quantum dynamics is to be described as shown in Fig.~\ref{F1a} and if
the time and the dynamics is to emerge through correlations between the
quantum clock and the system as described through an analogy with a movie
reel, a question then arises about the speed of such an unfolding of events
(states). Basically, it is the Hamiltonian which drives the time development
of the system (as is the case in the Schr\"{o}dinger equation). This is one
strong reason for relating the clock to the system Hamiltonian, and done in a
self-contained fashion by the Hamiltonian constraint in Eq.~(\ref{q3a}).
To illustrate this and show that a hypothetical external observer would
see the universe as static whereas an internal observer will see it developing
with time Moreva {\em et al.}~\cite{Mor} suggested an experimental set up using an
entangled polarization state of two photons with one photon being used as a
clock to measure the evolution of the second. They have shown that in their
experiment an internal observer that becomes correlated with the photon taken
as the clock will see the other photon evolve. However an external observer
who only observes the global properties of the two photons sees it as static.
Their scheme is an elegant illustration of the PW mechanism.

One important feature of quantum clocks arises once we have several equivalent
quantum clocks measuring time. Due to the multiple partitionings of the
Hilbert space, we would expect that time simultaneity is lost. This resembles
what happens in the theory of special relativity; different observers measure
different times of occurrence for the same event. We therefore need to
establish a metric for the quantum measurement of time, or perhaps we need
not. This remains an open question. Some proposals for quantum time in an
interacting clock-system formulation are available (see for example
\cite{smith}); however, the ambiguity of the disturbance caused by such an
interacting model remains questionable. As it seems, the
relatival description of the relation between the frames of reference and the
question of perspectives, which is bound to satisfy certain symmetry
requirements, is usually discussed in the context of the relativity of quantum
states \cite{Lov,Gia2,Van}. Specifically, a basic and straight forward
description is given by Giacomini \cite{Gia} in which the notion of the
quantum reference frame (QRF) associated with a quantum particle is used. In
this proposal the proper time of the particle is taken as a QRF and the
evolving parameters of the rest of the quantum system under consideration
coincides with it. Such QRF allows us to treat space and time on equal
footing, safeguarding the Lorentz covariance of space and time measurements
since we are dealing with the proper time. However, if we go back to the above
formulations of quantum time, which are based on the PW scheme, we can see
that this QFR is already built in the mechanism and is represented by the
quantum clock. The author correctly remarked on this connection between their
construction and the PW mechanism for non-interacting clocks when the external
degrees of freedom are neglected.

PW \cite{PW} have remarked that because an infinite ensemble is needed to
determine conditional probabilities, no prediction of quantum physics can ever
be completely verified by quantum-mechanical observers within the universe,
for whom the theory can only make statistical predictions. This may sound
discouraging to some extent since it implies that the observational
verification of the quantum measurement of time may not be practically
tenable. However, certainly there could be other approaches for chasing the
detection of quantum time effects. One example of the precise measurement of
quantum time effects could be through using the photometric effects of phase
shifts \cite{Zyc}.

At this stage it seems that the formulation of a consistent scheme of quantum
time has already reached an advanced stage and has already overcome many
fundamental difficulties. The formulations presented above are consistent and
beautiful. However, the question whether the internal time measured by quantum
clocks is affirming the philosophical choice of relatival time (that needs
changes to exist) or substantival time (that does not need changes to exist)
remains an open question \cite{Serg}. \\[0.5cm]
{\em Acknowledgement.}
M. B. A. would like to thank Alessandro Coppo, Tommaso Favalli, Flaminia
Giacomini, Philipp A. H{\"o}hn, Lorenzo Maccone, Alexander Smith, 
Ovidiu Cristinel Stoica and Vlatko Vedral for useful discussions. Tommaso Favalli provided
Fig.~\ref{F2} shown in the text. D. H. acknowledges financial support form the UK Engineering and Physical Sciences Research Council EPSRC (Award Ref.~Nr.~2130171).

\end{document}